\documentclass[runningheads]{llncs}
\usepackage[T1]{fontenc}
\usepackage{graphicx}
\usepackage{multirow}
\usepackage{amsmath}
\usepackage{url}
\begin{document}
\pagestyle{empty}
\title{An Optimized Pipeline for Automatic Educational Knowledge Graph Construction}
\titlerunning{An Optimized Pipeline for Automatic EduKG Construction}
%
\author{
Qurat Ul Ain\inst{1}\thanks{Email: \email{qurat.ain@stud.uni-due.de}}\orcidID{0000-0003-2691-0267} \and
Mohamed Amine Chatti\inst{1}\orcidID{0000-0002-1311-7852} \and
Jean Qussa\inst{1}\orcidID{0009-0003-3006-1111} \and
Amr Shakhshir\inst{1}\orcidID{0009-0005-4278-9681} \and
Rawaa Alatrash\inst{1}\orcidID{0000-0003-2192-028X} \and
Shoeb Joarder\inst{1}\orcidID{0000-0003-4591-9742}
}

\authorrunning{Ain et al.}

\institute{
Social Computing Group, Faculty of Computer Science, University of Duisburg-Essen, Germany
}
\maketitle              
\begin{abstract}
The automatic construction of Educational Knowledge Graphs (EduKGs) is essential for domain knowledge modeling by extracting meaningful representations from learning materials. Despite growing interest, identifying a scalable and reliable approach for automatic EduKG generation remains a challenge. In an attempt to develop a unified and robust pipeline for automatic EduKG construction, in this study we propose a pipeline for automatic EduKG construction from PDF learning materials. The process begins with generating slide-level EduKGs from individual pages/slides, which are then merged to form a comprehensive EduKG representing the entire learning material. We evaluate the accuracy of the EduKG generated from the proposed pipeline in our MOOC platform, CourseMapper. The observed accuracy, while indicative of partial success, is relatively low particularly in the educational context, where the reliability of knowledge representations is critical for supporting meaningful learning. To address this, we introduce targeted optimizations across multiple pipeline components. The optimized pipeline achieves a 17.5\% improvement in accuracy and a tenfold increase in processing efficiency. Our approach offers a holistic, scalable and end-to-end pipeline for automatic EduKG construction, adaptable to diverse educational contexts, and supports improved semantic representation of learning content.
%
\end{abstract}
\section{Introduction}
Knowledge graphs (KGs) have gained significant attention in recent years due to their ability to structure and integrate diverse information across various domains \cite{hogan2021knowledge}. 
Educational knowledge graphs (EduKGs) refer to the application of KGs in the
educational domain, structuring knowledge concepts and their relationships to support personalized learning and adaptive education \cite{info14100526}. They play a crucial role in technology-enhanced learning (TEL) systems such as massive open online courses (MOOCs) and learning management systems (LMSs) by extracting and visualizing key learning concepts \cite{qu2024survey}. Structurally, an EduKG is a heterogeneous graph where nodes represent educational entities (e.g., knowledge concepts, learning materials) and edges define various types of semantic relationships between them \cite{qu2024survey}.
EduKGs serve a wide range of purposes in the educational domain, including enhancing learning and scientific discovery \cite{chen2018knowedu}, educational assessment \cite{fettach2022knowledge}, training KG practitioners \cite{engelbrecht2024teaching}, learner modeling \cite{umaprawaa}, personalized learning resource recommendation \cite{QURATlak}, and so on. 

With the exponential growth of digital learning content, there is an increasing need for automated and scalable methods to organize and interpret domain knowledge effectively. EduKGs offer a robust framework for modeling such knowledge, yet the automatic construction of high-quality EduKGs remains a significant challenge. 
Traditionally, EduKGs were constructed manually by domain experts. However, due to the time and effort required for this approach, recent research has shifted towards automating EduKG generation using machine learning (ML) and natural language processing (NLP) techniques \cite{2023surveykgs}. 
In parallel, the emergence of Generative AI and Large Language Models (LLMs) has sparked interest in their potential for EduKG generation \cite{jhajj2024educational,zhu2024llms}. Despite their promise, LLMs face significant challenges in automatic EduKG generation, such as token limits that restrict processing large datasets, inefficient human-machine interaction that either lacks oversight or increases labor costs, and a tendency to generate inaccurate information due to their hallucinating nature. These limitations make LLMs unreliable for fully autonomous EduKG construction \cite{zhu2024llms}.
Despite growing research interest in automatic EduKG construction, there remains a pressing need for a holistic, reliable, and scalable end-to-end pipeline that can automatically generate high-quality EduKGs across diverse educational settings. Existing efforts such as EduKG for K-12 \cite{zhao2022edukg} are ontology-driven, requiring a predefined schema before extraction rather than fully automatic extraction from unstructured PDFs.  
As part of our ongoing research efforts to address this challenge \cite{info14100526,ain2025top}, this study focuses on refining and optimizing the automatic EduKG construction process to develop a unified and robust solution. We first propose a pipeline for automatic EduKG construction that can process any PDF-based learning material,
starting from a single page/slide of a PDF, leading to a slide-EduKG per slide which are then combined to form an EduKG to represent the whole learning material. We then evaluate the accuracy of the EduKG generated by the proposed pipeline in our MOOC platform, CourseMapper \cite{ain2022learning}. The accuracy of the EduKG was not very high which calls for a need for further improvements in the pipeline. Subsequently, we introduce a series of optimizations, including enhancements to the pipeline architecture, improved data preprocessing, and refined modules for text extraction, concept annotation and EduKG expansion. The optimized pipeline is then compared against its initial version in terms of accuracy and efficiency, achieving a 17.5\% improvement in accuracy and a tenfold increase in processing efficiency.
All experiments were conducted within CourseMapper, to demonstrate the pipeline’s applicability in real-world educational environments. The resulting pipeline offers a state-of-the-art, domain-independent solution for automatic EduKG construction and can be easily adapted to a wide range of educational contexts.
\section{EduKG Construction Pipeline} \label{pipelineoverview}
The EduKG construction pipeline consists of several steps, as described below.
\subsection{Text Extraction}\label{textext}
The first step in building EduKG involves extracting text from individual pages/ slides of PDF learning materials. This step is particularly challenging due to the diverse and often inconsistent layouts and structures of educational PDFs. To ensure coherent input for the subsequent keyphrase extraction step, it is essential to accurately identify and reconstruct logically connected text segments in the correct reading order. We address this using PDFMiner \cite{ramakrishnan2012layout} which applies a layout-aware strategy by detecting text blocks, applying rule-based classification, and merging them to preserve the intended textual sequence.
\subsection{Keyphrase Extraction}
Once text is extracted from a page/slide of the PDF learning material, keyphrase extraction is performed as a preliminary step to Wikipedia entity linking, optimizing the process by minimizing the amount of text processed by the entity linking service. To determine the most effective keyphrase extraction method, we evaluate both state-of-the-art Large Language Model (LLM) approaches including Zero-Shot, Few-Shot, and KeyBERT-assisted prompts, and traditional techniques, including SingleRank, PatternRank, and SIFRank variants.
The Inspec \cite{inspec} and SemEval \cite{semeval} datasets serve as benchmarks for performance evaluation, as they contain scholarly texts relevant to the educational domain, making them suitable for this evaluation.
Our experiments apply the LLM-based methods using the Mistral-7B-Instruct-v0.1 model \cite{mistral}, while the traditional methods are implemented following their respective original configurations. Mistral-7B, an open-source instruction-tuned LLM with 7 billion parameters, is well-suited for this task due to its strong language understanding, ability to follow prompts, and scalability for large datasets. 
In the Zero-Shot setting, the LLM extracts keyphrases based solely on its pre-trained knowledge, using prompts such as “Identify key phrases from the following text: [Document Text]”. The Few-Shot setting improves performance by including a few example text–keyphrase pairs in the prompt. The LLM-KeyBERT approach combines LLM prompting with KeyBERT \cite{keybert}, which uses BERT embeddings to extract contextually relevant keywords, enhanced by the LLM’s nuanced understanding of the input. All methods are evaluated by comparing the top-n extracted keyphrases against expert annotations in the Inspec and SemEval datasets. Effectiveness is measured using Precision, Recall, and F1-score, capturing each method’s accuracy, coverage, and balance. The results are presented in Table \ref{tab:keyphrase}. This comparative analysis shows that LLM-based approaches, particularly with Few-Shot learning, outperform others in keyphrase extraction. This highlights the significance of LLMs' nuanced understanding and adaptability in capturing semantically rich keyphrases. While traditional methods lag behind in performance, they still hold significant value, especially in contexts where access to advanced models is limited. It is worth noting that the reported results for LLM-based methods reflect their best-case performance; in practice, some runs show a 10–15\% drop in effectiveness, emphasizing the need for careful validation when deploying these models.
\begin{table}
\centering
\caption{A comparison of different keyphrase extraction methods}
\label{tab:keyphrase}
\begin{tabular}{|c|l|c|c|c|c|c|c|}
\hline
K & Keyphrase Extraction Method & \multicolumn{3}{c|}{Inspec Dataset} & \multicolumn{3}{c|}{SemEval Dataset} \\ \hline
\multicolumn{2}{|c|}{} & P & R & F1 & P & R & F1 \\ \hline
\multirow{7}{*}{5} & Zero-shot LLM & 72.95 & 75.17 & 74.04 & 72.95 & 75.17 & 73.2 \\
 & Few-shot LLM & \textbf{74.28} & \textbf{76.14} & \textbf{75.20} & \textbf{74.28} & \textbf{76.14} & \textbf{76.33} \\
 & LLM\_KeyBERT & 71.51 & 73.81 & 72.64 & 71.51 & 73.81 & 72.12 \\
 & SingleRank & 64.78 & 67.04 & 67.86 & 64.75 & 67.04 & 67.15 \\
 & PatternRank & 60.83 & 62.98 & 61.89 & 60.83 & 62.98 & 61.45 \\
 & SIFRank & 65.21 & 67.49 & 66.33 & 65.21 & 67.49 & 68.11 \\
 & \begin{math}SIFRank_{SqueezeBERT}\end{math} & 66.75 & 67.04 & 65.89 & 66.75 & 69.01 & 66.24 \\ \hline
\multirow{7}{*}{10} & Zero-shot LLM & 69.14 & \textbf{84.98} & 76.25 & 69.14 & 84.98 & 75.6 \\
 & Few-shot LLM & \textbf{72.04} & 84.57 & \textbf{77.81} & \textbf{72.04} & \textbf{84.57} & \textbf{78.18} \\
 & LLM\_KeyBERT & 66.79 & 83.00 & 74.02 & 66.79 & 83.00 & 74.31 \\
 & SingleRank & 66.11 & 81.19 & 73.09 & 66.11 & 81.19 & 73.23 \\
 & PatternRank & 64.11 & 79.02 & 70.79 & 64.11 & 79.02 & 70.91 \\
 & SIFRank & 62.74 & 77.06 & 69.16 & 62.74 & 77.06 & 69.23 \\
 & \begin{math}SIFRank_{SqueezeBERT}\end{math} & 66.34 & 81.37 & 72.87 & 66.34 & 81.37 & 73.86 \\ \hline
\multirow{7}{*}{15} & Zero-shot LLM & 59.16 & \textbf{89.08} & 71.10 & 59.16 & \textbf{89.08} & 71.2 \\
 & Few-shot LLM & \textbf{66.61} & 88.15 & \textbf{75.88} & \textbf{66.61} & 88.15 & \textbf{75.68} \\
 & LLM\_KeyBERT & 55.65 & 87.27 & 67.96 & 55.65 & 87.27 & 68.16 \\
 & SingleRank & 57.12 & 87.02 & 69.01 & 57.12 & 87.00 & 69.12 \\
 & PatternRank & 55.86 & 85.98 & 67.72 & 55.86 & 85.98 & 68.09 \\
 & SIFRank & 54.49 & 83.05 & 65.81 & 54.49 & 83.05 & 66.21 \\
 & \begin{math}SIFRank_{SqueezeBERT}\end{math} & 57.17 & 87.00 & 68.96 & 57.17 & 87.00 & 69.33 \\ \hline
\end{tabular}
\end{table}
Another important dimension of our evaluation is runtime efficiency. LLM-based methods, although accurate, demand significantly more time and computational resources per document than traditional methods (see Table \ref{tab:time}). Despite lacking the depth of LLM-based analysis, traditional approaches offer dependable and efficient solutions, particularly when computational constraints or real-time processing are priorities. Among these, \begin{math}SIFRank_{SqueezeBERT}\end{math} demonstrates a strong balance between accuracy and efficiency, making it a practical choice for keyphrase extraction in our pipeline. The keyphrase extractor extracts n=15 keyphrases from each slide of the PDF material, because in a learning environment it would be overwhelming for
the learners to focus on more than 15 concepts within a slide.
\begin{table}
    \centering
\caption{Runtime required to extract (N=5) keyphrases per document}
\label{tab:time}
    \begin{tabular}{|c|c|} \hline
         Keyphrase extraction method& Runtime (min:sec)\\ \hline
         LLM-based methods& 1:40\\
         SingleRank& <0:01\\
         PatternRank& <0:01\\
         SIFRank& 0:03\\
         \begin{math}SIFRank_{SqueezeBERT}\end{math}& 0:24\\
\hline    \end{tabular}
\end{table}
\subsection{Concept Identification} \label{CI}
The extracted keyphrases are used for concept identification by linking them to external knowledge bases. Following \cite{Manrique2018knowledge}, we use DBpedia Spotlight \cite{mendes2011dbpedia} to map keyphrases to DBpedia concepts through spotting, candidate selection, and disambiguation based on their contextual information. The annotated keyphrases are then referred to as Main Concepts (MCs) of the slide. Nevertheless, entity-linking tools like DBpedia Spotlight can occasionally produce incorrect annotations where keyphrases are linked to irrelevant concepts, due to its fully automated nature \cite{Manrique2018knowledge}. To address this issue, we implement a concept-weighting strategy that emphasizes semantic and contextual relevance. We use a transformer-based method ($w_{SBERT}$), where weights of the MCs ($w_{LM}$) are computed via cosine similarity between SBERT \cite{reimers2019sentence} embeddings of the whole learning material and corresponding Wikipedia articles. Additionally, the similarity score between the embedding of the text of the slide and the embedding of the Wikipedia abstract of the concept is obtained, resulting in a slide similarity score ($w_{Slide}$). The relevance of the concepts per slide is determined by using the sum of these two similarity scores (($w_{LM}$) and ($w_{Slide}$)). After obtaining the slide’s MCs and their weights, relations between MCs and the corresponding slide, as well as relations between MCs and the learning material, are established to ensure that concepts exist at both the learning material and slide levels.
At this point, the Slide-EduKG containing the current slide, MCs, learning material, and their relations are stored in a Neo4j database and made available for learners, even if the construction process has not yet been completed. By storing the data after annotating each slide, allows learners to explore Slide-EduKGs for completed slides without waiting for the full EduKG construction to finish. Additionally, combining slide-level and learning material-level weights ensures balanced concept importance, reducing bias toward either the slide or the overall learning material. Once all slides have been annotated, the \textit{concept expansion} step is activated.
\subsection{Concept Expansion} \label{expansionold}
Since keyphrase extraction and concept identification may overlook some relevant concepts \cite{manrique2018weighting}, we expand the identified concepts to increase EduKG coverage and diversity to better support knowledge exploration. To achieve this, we apply two expansion strategies based on semantic relationships in DBpedia, as proposed by Manrique et al. \cite{Manrique2018knowledge}: related concept expansion and category-based expansion.
In related concept expansion, semantically linked concepts are added to the EduKG using SPARQL queries over the dbo:wikiPageWikiLink property. Whereas, in category-based expansion, hierarchical categories are retrieved via the dct:subject property. 
After identifying the related concepts (RCs) and categories (Cts), their embeddings are generated using SBERT. For RCs, the embeddings are derived from the abstracts of their corresponding Wikipedia articles, whereas Cts' embeddings are based on their names. Cosine similarity is then calculated between the embeddings of each MC and its associated RCs and Cts to assess their semantic closeness. Additionally, cosine similarity is measured between the RCs and Cts and the whole learning material. These two similarity scores are summed up to compute a unified weight for each RC and Ct. This overall weight is used to rank them allowing the most relevant RCs and Cts to be selected. Specifically, the top 20 RCs and top 3 Cts for each MC are retained and integrated into the EduKG. This method enhances the structure of the EduKG by incorporating and exploring semantically related concepts and categories.

After building the EduKG slide by slide, it is then aggregated to create a LM-EduKG, that is a composite of all the slide-EduKGs. This approach enforces two key constraints: 1) The concepts related to a slide must adequately capture the content of that slide, and, 2) Any concept appearing at the lower level i.e., Slide-EduKG should also be included at the higher level i.e., LM-EduKG. 
\section{Evaluation}\label{firsteval}
For the evaluation we used CourseMapper, to construct and visualize EduKG from PDF learning materials using the proposed pipeline. 
We evaluate the accuracy of the generated EduKG using the simple random sampling (SRS) method proposed by Gao et al. \cite{Gao2019}. This method evaluates the correctness of samples in a KG, which consist of triples of (subject, predicate, object), where subject and object can be any two nodes in the KG, and predicate is the relation that links them. The evaluation involves two key tasks: entity identification which requires the evaluators to verify the correctness of a node entity by considering contextual information, and relationship validation which assesses the correctness of relations between subject and object nodes. 
The final accuracy score \begin{math} \mu_s \end{math} is computed as the mean of sample judgments. To ensure statistical reliability, the method follows an iterative sampling process with a predefined margin of error threshold. If the margin of error for the initial sample set exceeds the threshold, additional samples are evaluated until the accuracy stabilizes within the acceptable range.
The results are presented using the normal approximation, which is the mean value \begin{math} \pm \end{math} the standard deviation value \begin{math} \sigma \end{math}. The mean value of the evaluated sample set \begin{math} \mu_s \end{math} represents accuracy and falls between 0 and 1, where higher values indicate greater accuracy. Two course experts of the same course were recruited and an EduKG was generated for evaluation using a random PDF material from their course in CourseMapper.
The evaluation process took 4 hours where the first annotator evaluated 200 random samples from the generated EduKG, while the second annotator evaluated 180 random samples, making sure that each annotator evaluated a different sample set.
The mean value of accuracy \begin{math} \pm
\end{math} standard deviation (\begin{math}\mu_s \pm \sigma \end{math}) is found to be 0.4 \begin{math}\pm \end{math} 0.049. This indicates that approximately 40\% of the evaluated triples were judged as correct with respect to both entity identification and relationship validation. The standard deviation of 0.049 indicates relatively low variability in the judgments across the sample, implying that the accuracy estimates are consistent and reliable within the evaluated subset. The narrow spread strengthens confidence in the stability of the reported accuracy and suggests that the results are representative of the overall quality of the EduKG.
\section{Pipeline Optimization}\label{newpipe}
The observed accuracy of EduKG generated by the proposed pipeline, while indicative of partial success, is relatively low particularly in the educational context, where the reliability of knowledge representations is critical for supporting meaningful learning. This underscores the need for systematic optimizations within the pipeline to ensure the EduKGs are more accurate, efficient, and easily accessible, to support effective learning experiences. In an effort to address these limitations, we propose a series of optimizations across various steps of the initial pipeline, with the anticipation of achieving higher accuracy and efficiency.
\subsection{Worker-based Architecture}
One major limitation of the proposed pipeline is its implementation as an API within a Python-Flask application. While generally a reasonable approach, it is unsuitable for the time-intensive EduKG generation process. Handling multiple simultaneous requests can overload the system, especially when users request the EduKGs for multiple materials at once. Additionally, failures at any stage leave the database in an inconsistent state with no straightforward recovery method. The current system also lacks request queuing, forcing users to retry if the server is busy.
To address these issues, a worker-based architecture is proposed. This approach utilizes job queues and multiple workers that continuously process queued EduKG construction tasks. Each worker retrieves a job, executes it, stores the results in the database, and updates the user about its status. This architecture enables horizontal scalability by adding more workers as needed, ensures failed jobs are automatically re-queued, and allows request queuing when all workers are busy. Redis is chosen as the message broker due to its simplicity, reliability, and efficiency as an in-memory database.
\subsection{Data Preprocessing} \label{wiki}
A major bottleneck in the proposed pipeline is its reliance on the Wikipedia API particularly in concept annotation, weighting and expansion modules, which introduces significant latency due to a huge number of network requests. Additionally, generating embeddings for identified concepts further impacts performance. To mitigate these issues, an offline data preprocessing step is proposed. This step, executed once a month, extracts relevant data from the Wikipedia XML dump, including article names, abstracts, inter-article links, category associations, redirections, and disambiguation pages. The extracted data is then used to generate embeddings for concept abstracts and category names using the SBERT all-mpnet-base-v2 sentence embedding model. The processed information is stored in a PostgreSQL database, allowing near-instant access to embeddings, categories, and relationships during EduKG generation.
While this approach increases storage requirements and preprocessing time, the benefits of eliminating network latency and reducing redundant computations significantly outweigh these costs. Moreover, the database remains accessible to EduKG pipeline workers even if Wikipedia API is down or being updated.
\subsection{Optimization of Text Extraction}
The current pipeline originally employed a naive text extraction method using PDFMiner. However, this approach lacked any preprocessing, resulting in significant inaccuracies caused by the inclusion of noisy elements such as slide titles, page numbers, references, and footers. To improve the reliability of extracted content, the text extraction module is substantially enhanced to include structured analysis aimed at determining the relevance of extracted elements and ensuring accurate segmentation of sentences and bullet points. The improved approach incorporates several analytical techniques to ensure cleaner and more meaningful output. \textit{Font size analysis} is used to assess whether adjacent text elements should be treated as part of the same sentence. If the difference in font size between two elements exceeds 0.5 points, they are considered separate sentences. In addition, \textit{text distance analysis} is applied to account for vertical and horizontal spacing; when the distance between two adjacent elements is more than 1.5 times the respective line height or width, segmentation occurs at that point. To filter out repetitive or irrelevant content such as footers, sidebars, and page numbers, \textit{text similarity analysis} is also employed. After removing non-alphabetical characters, any text fragment that appears in the same location on more than half of the pages, or on at least five pages in documents with fewer than ten pages, is classified as noise and removed. Furthermore, \textit{bullet point analysis} is integrated to correctly segment list items, addressing inconsistencies introduced by various PDF rendering tools. Bullet points embedded within text naturally serve as sentence separators. However, when bullet points are rendered as separate graphical elements, such as circles or rectangles, additional spatial analysis is required. These elements are identified based on their position relative to adjacent text, under the assumption of a left-to-right writing system. Once a bullet point is detected, the following text is treated as the start of a new sentence. Overall, this enhanced methodology substantially improves the accuracy of the text extraction module and consequently keyphrase identification by filtering irrelevant content and preserving the semantic structure of the learning material.
\subsubsection{Evaluation and Results:}
To evaluate the impact of the proposed changes, we compared the text extracted from the optimized module with the initial one (see Section \ref{textext}). Existing datasets like Grotoap2 \cite{tkaczyk2014grotoap2}, PubMed and BioMed Central \cite{bmc_2018_bmc} etc., mainly contain text from scientific articles, which do not align with our use case of extracting text from slides and PDF learning materials. Therefore, a different evaluation approach is required. Bast et al. \cite{bast_2017_a} recommend extracting structure from TeX files and evaluating text extraction based on the rendered PDFs, but this method is applicable only to LaTeX-based presentations, which are a small subset of all learning materials. For this evaluation, a new dataset was created with materials from various sources i.e. LaTeX, Microsoft PowerPoint, and Microsoft Word. A total of 24 PDF materials were selected, ensuring variety in layout and content types.
The evaluation consisted of three steps. First, text was extracted from the slides using PDFMiner. Second, a human annotator labeled each text block with categories: header, content, footer, page number, navigation, figure, or empty.  The final dataset contained 643 annotated text parts from 24 slides across 7 different learning materials. Third, comparing the human annotations to the output of both text extraction modules and calculate precision, recall and F1 scores. We further utilized the text extraction evaluation criteria proposed by Bast et al. \cite{bast_2017_a}, which is divided into three groups: newline differences, which capture the quality of the detection of paragraph boundaries ($NL^{+}$ is the number of spurious newlines in the output file, whereas $NL^{-}$ is the number of missing newlines in the output file); paragraph differences, which capture the quality of the distinction between body and non-body text and of the reading order ($P^{+}$ is the number of spurious paragraphs, $P^{-}$ is the number of missing paragraphs, and $P^{\updownarrow}$ is the number of rearranged paragraphs in the output file); and word differences, which capture the quality of the recognition of individual words and their boundaries ($W^{+}$ is the number of spurious words, $W^{-}$ is the number of missing words, and $W^{\sim}$ is the number of misspelled words in the output file). The results are shown in Table \ref{tab:text} and \ref{tab:text2}, which clearly demonstrate that our new optimizations outperform the earlier one in nearly every measure. 

\begin{table}
\centering
\caption{Comparison of the optimized text extraction vs. earlier one}
\label{tab:text}
\begin{tabular}{|c|c|c|c|c|}
\hline
Text Extraction Type                                                                        & Module Type & P    & R    & F1   \\ \hline
\multirow{2}{*}{Content and figure}                                                         & Initial         & 0.64 & 0.96 & 0.77 \\
                                                                                            & Optimized         & \textbf{1.0}  & \textbf{0.97} & \textbf{0.98} \\
\multirow{2}{*}{Non-content}                                                                & Initial         & 0.0  & 0.0  & 0.0  \\
                                                                                            & Optimized         & \textbf{0.93} & \textbf{1.0}  & \textbf{0.96} \\
\multirow{2}{*}{\begin{tabular}[c]{@{}l@{}}Beginning of sentence in content\end{tabular}} & Initial         & 0.82 & 0.08 & 0.14 \\
                                                                                            & Optimized         & \textbf{0.94} & \textbf{0.95} & \textbf{0.95} \\
\multirow{2}{*}{\begin{tabular}[c]{@{}l@{}}Beginning of sentence in figure\end{tabular}}  & Initial         & 0.0  & 0.0  & 0.0  \\
                                                                                            & Optimized         & \textbf{0.36} & \textbf{0.38} & \textbf{0.37}
\\
\hline
\end{tabular}
\end{table}
\begin{table}
\centering
\caption{Results of the evaluation of the initial vs. optimized text extraction using the differences criteria proposed by \cite{bast_2017_a}}
\label{tab:text2}
\begin{tabular}{|c|c|c|c|c|c|c|c|c|}
\hline
Module & $NL^{+}$ & $NL^{-}$ & $P^{+}$ & $P^{-}$ & $P^{\updownarrow}$ & \textbf{$W^{+}$} & $W^{-}$ & $W^{\sim}$ \\ \hline
	Initial & \textbf{3} & 230 & 230 & \textbf{12} & 1 & 0 & 0 & 1 \\ \hline
	Optimized & 52 & \textbf{49} & \textbf{1} & 17 & 1 & 0 & 0 & 1 \\ \hline
\end{tabular}
\end{table}
\subsection{Optimization of Concept Annotation}
In the concept annotation module, extracted keyphrases are sent to the DBpedia Spotlight API for annotation. Multiple requests are processed simultaneously, making the total annotation time dependent on the longest individual request, aside from network overhead. 
However, this approach has three main limitations: (1) DBpedia Spotlight may return irrelevant or miss relevant annotations, (2) sending only keyphrases omits contextual information crucial for accurate interpretation, and (3) as an online service, DBpedia Spotlight can be inconsistent and unreliable.
To mitigate these issues, a \textit{concept weighting} and a \textit{concept disambiguation} modules are integrated to filter irrelevant concepts and enhance annotation accuracy in the EduKG.
Earlier we determined the weight of each MC with respect to each slide and to the whole learning material (see Section \ref{CI}). We use this same strategy (named as concept weighting) further in concept disambiguation; a process that enhances EduKG accuracy by identifying better alternatives to DBpedia-annotated concepts. It checks whether an annotated concept is the title of or linked from a Wikipedia disambiguation page. A disambiguation page is a Wikipedia page that lists articles sharing the same or a similar title. If this condition is met, all linked pages are considered alternative concepts, which are then weighted by the concept weighting strategy. The highest-weighted concept replaces the original annotation.
This step helps improve the quality of the EduKG by providing alternative annotations for keyphrases that may have been incorrectly annotated in the concept annotation step. This is because DBpedia only knows the keyphrase that is passed, but not the context thereof.
The next step is \textit{knowledge graph pruning}, i.e., removing irrelevant concepts from the EduKG. Irrelevant concepts are defined as concepts with a low weight as determined by the weighting strategy. All concepts whose weight falls below the weight threshold i.e. 0.192 (based on experiments mentioned below), are removed from the EduKG. 
\subsubsection{Evaluation and Results:}
The performance of this module is influenced by three parameters: the sentence embedding model, the weight threshold, and the inclusion of the disambiguation step.
The sentence embedding model, used to compute concept-material similarity, significantly impacts concept weighting. 
Three pretrained transformer models are compared: all-mpnet-base-v2 \cite{pretrained}, known for its accuracy but slow performance; all-MiniLM-L6-v2 \cite{pretrained}, offering a balance between speed and accuracy; and msmarco-distilbert-base-tas-b \cite{huggingface2023msmarco}, optimized for retrieval tasks. The evaluation uses the CCI dataset \cite{Manrique2018knowledge}, consisting of transcripts from 96 programming-related learning videos annotated by seven experts. The weight threshold is ignored, and instead, the top-1, top-2, top-3, top-5, and top-10 concepts per resource are selected and evaluated using Precision (see Table \ref{tab:precision_models}). Results show that all-mpnet-base-v2 matches expert-annotated concepts highest, followed closely by all-MiniLM-L6-v2 and msmarco-distilbert-base-tas-b, which perform similarly. Due to its superior performance, all-mpnet-base-v2 is selected to be used for embeddings generation.
\begin{table}[!h]
\centering
    \caption{Precision of the sentence embedding models}
    \label{tab:precision_models}
    \begin{tabular}{|c|c|c|c|c|c|} \hline
        Model & Top-1 & Top-2 & Top-3 & Top-5 & Top-10 \\ \hline
        msmarco-distilbert & 0.18 & 0.16 & 0.14 & 0.10 & 0.07 \\
        all-MiniLM-L6-v2 & 0.17 & 0.15 & 0.13 & 0.09 & 0.06 \\
        all-mpnet-base-v2 & \textbf{0.25} & \textbf{0.19} & \textbf{0.16} & \textbf{0.13} & \textbf{0.09} \\ \hline
    \end{tabular} 
\end{table}

An additional experiment was conducted to determine the optimal weight threshold for knowledge graph pruning. Using the CCI dataset \cite{Manrique2018knowledge}, expert-annotated concepts were weighted against transcripts using our weighting strategy. Results show that 50\% of concepts have a weight above 0.471, where weight falls in the range of $[-1, 1]$ with all the weights falling between 0.028 and 0.744.
However, using such a low threshold would introduce excessive noise into EduKG. Since 95\% of weights fall between 0.192 and 0.774, 0.192 is selected as the weight threshold, meaning that all concepts having a weight less than 0.192 will be removed from the EduKG (knowledge graph pruning step).

Lastly, we evaluated the impact of introducing disambiguation step on the generated EduKG accuracy. Using the CCI dataset \cite{Manrique2018knowledge}, EduKG was generated twice, once with and once without disambiguation. Precision, Recall, and F1-score were computed (see Table \ref{tab:disambiguation_results}).
Results confirm that disambiguation enhances EduKG accuracy, emphasizing the importance of considering context rather than just keyphrases during annotation. Consequently, the disambiguation step is included in the optimized EduKG construction pipeline.
\begin{table}
    \centering
        \caption{Concept annotation with and without concept disambiguation}
    \label{tab:disambiguation_results}
    \begin{tabular}{|c|c|c|c|}
    \hline
      Method  & Precision & Recall & F1-Score \\ \hline
        Initial Pipeline (Without Disambiguation) & 0.13 & 0.17 & 0.14 \\
        Optimized Pipeline (With Disambiguation) & \textbf{0.18} & \textbf{0.29} & \textbf{0.23} \\ \hline
    \end{tabular}
\end{table}
\subsection{Optimization of Concept Expansion}
Concept expansion is the process of assigning related concepts (RCs) and Categories (Cts) to the identified main concepts (MCs). In the proposed initial pipeline, these RCs and Cts were retrieved from DBpedia using SPARQL queries. While SPARQL queries offer a straightforward method for retrieving RCs and Cts, they come with practical limitations. Specifically, querying public SPARQL endpoints at scale can lead to performance bottlenecks due to latency, rate limiting, or inconsistent response times, making the approach less suitable for high-throughput or time-sensitive applications. Moreover, relying on external infrastructure introduces risks related to availability and long-term stability, as these endpoints may become temporarily inaccessible or deprecated without warning. To address these issues, we instead retrieve RCs and Cts from our locally hosted Wikipedia dump (Section \ref{wiki}). This alternative ensures faster and more reliable access to the required data while giving us full control over the extraction process. It also eliminates external dependencies, enabling more scalable and robust integration within our pipeline.

Related-concept expansion consists of three steps: candidate set creation, candidate set weighting, and candidate pruning. It works by creating a set of expansion candidates for each MC. This expansion candidate set contains all Wikipedia pages linking and linked to from the Wikipedia page of the MC. These candidates are then weighted by computing the cosine similarity between the embedding of the abstract of the Wikipedia page of the candidate and the embedding of the text of the whole PDF material. After every candidate RC has been assigned a weight, they are sorted by weight and the top-20 candidates for each MC are chosen as the final set of RCs for that MC.
Similarly, category expansion process consists of candidate set creation, candidate set weighting, and candidate pruning. 
The category expansion candidate set consists of all categories of the all the MCs of the PDF material. The categories of the concepts are the Wikipedia categories to which the Wikipedia article of the concept belongs. This list of categories is retrieved from the Wikipedia dump database which was created
in the preprocessing step (Section \ref{wiki}). The candidate set is then weighted by computing the product of the normalized category weight
and the connected concepts weight. The normalized category weight ($w_{nc}$) is the cosine similarity between the embedding of the category name and the embedding of the text of whole PDF material. The connected concepts weight ($w_{cc}$) is the inverse of the logarithm of the number of MCs connected to the category plus one. The one is added to avoid zero weights. Using the number of connected MCs to a category allows more relevant categories to the learning material, being assigned a higher weight. The embedding of the category name is precomputed and is simply retrieved from the database.
Based on this final weight, the top-5 categories for each MC are then kept and the rest are discarded.
\section{Evaluation of the Optimized Pipeline}\label{evalnew}
\label{sec:newevaluation}
The optimizations proposed in various steps of the EduKG construction pipeline were evaluated in terms of efficiency and accuracy of the generated EduKG.
\subsection{Evaluation of EduKG Efficiency}
The runtime of the optimized pipeline was evaluated and compared to the initial version of the pipeline on a machine with the following specifications: CPU: AMD Ryzen 5 5600, RAM: 32GB, Storage: 4TB NVMe M.2 SSD, GPU: None. The GPU was intentionally excluded to reflect performance on a typical commodity server.
The speed and memory usage of the pipeline was measured using the following metrics: 1) Mean time per slide, 
computed as the time required to extract MCs from a PDF learning material divided by the number of slides in the material, and 2) Mean time per concept,
computed as the time required to expand the knowledge graph in a learning material divided by the number of MCs in the material. Furthermore, EduKG creation and expansion were evaluated separately. A diverse dataset of learning materials, varying in size, structure, density, and topics, was created. To test scalability, the dataset included one large material with 279 pages, while the remaining materials had 31 to 52 pages. The initial and optimized version of the pipeline were run on each material in the dataset, with DBpedia response caching disabled to ensure results accurately reflect performance on new, unseen data.

The evaluation results (see Table \ref{tab:evaluation-kg-overall}) show that the optimized pipeline is over 10 times faster in EduKG construction and more than 100 times faster in EduKG expansion. This substantial improvement in execution time compared to the initial version is primarily due to the use of a preprocessed local Wikipedia database with precomputed embeddings. Leveraging this local dump for both concept annotation and concept expansion significantly reduces computational overhead and enhances the overall reliability of the pipeline.
\subsection{Evaluation of EduKG Accuracy}
This evaluation aims to assess the accuracy of the EduKG generated with the proposed optimizations in the pipeline and compare it to the EduKG generated with the initial version of the pipeline (Section \ref{pipelineoverview}). The optimized pipeline is evaluated using the same approach as earlier, employing SRS and expert annotators. To ensure comparability, the same evaluation method, PDF material, metrics, and parameters are used. The process begins with randomly selecting a sample from the EduKG generated by the optimized pipeline, which is then annotated by the same two experts until the stopping criteria are met. 
The expert annotation part of the evaluation took around 2.5 hours in total. The first expert annotated a random set of 200 triples and the second one annotated 203 triples till we met the stopping criteria. The results of the evaluation are shown in Table
\ref{tab:evaluation-kg-overall}. 

The evaluation reveals a 17.5\% improvement in EduKG accuracy compared to the initial version of the pipeline, underscoring the effectiveness of the proposed optimizations. Particularly, improvements in the text extraction step played a critical role by significantly reducing noise in text, which previously led to incorrect or redundant annotations. Furthermore, the integration of concept disambiguation and knowledge graph pruning effectively filtered out irrelevant concepts, contributing to a more accurate and semantically coherent EduKG. The adoption of the all-mpnet-base-v2 embedding model further strengthened the pipeline, enabling the generation of high-quality embeddings that substantially improved the performance of the concept weighting step. Despite these advancements, the overall increase in accuracy remains moderate. This can be attributed to the continued dependence on DBpedia as the primary external knowledge base for concept annotation which is an element common to both versions of the pipeline. While the shared knowledge source may have limited the potential for drastic accuracy gains, the proposed optimizations nonetheless yielded meaningful improvements in both the accuracy and efficiency of EduKG construction, demonstrating the robustness and scalability of the optimized pipeline.
\begin{table}[h]
\centering
\caption{Evaluation results of EduKG accuracy and efficiency}
\label{tab:evaluation-kg-overall}
\begin{tabular}{|l|l|c|c|}
\hline
& \textbf{Metric} & \textbf{Initial Pipeline} & \textbf{Optimized Pipeline} \\ \hline
\multirow{2}{*}{Efficiency} 
  & Avg. Time per Slide (Generation) & 24.35 s & \textbf{2.3 s} \\
  & Avg. Time per Concept (Expansion) & 222 s & \textbf{1.89 s} \\ \hline
\multirow{2}{*}{Accuracy} 
  & Accuracy ($\mu_s$) & 0.4 & \textbf{0.47} \\
  & Accuracy ($\mu_s \pm \sigma$) & $0.4 \pm 0.049$ & \textbf{$0.47 \pm 0.049$} \\ \hline
\end{tabular}
\end{table}
\section{Conclusion and Future Work} \label{concl}
This study presents an end-to-end pipeline for the automatic construction of Educational Knowledge Graphs (EduKGs) from PDF learning materials within the MOOC platform, CourseMapper. The pipeline generates slide-level EduKGs from individual pages/slides of the PDF learning material and subsequently merges them into a comprehensive EduKG representing the entire learning material. Initial evaluation of the EduKG generated by the proposed pipeline indicated moderate performance, emphasizing the importance of further optimization to ensure that EduKGs accurately reflect educational content and support meaningful learning experiences.
To address this, we proposed several optimizations across multiple steps of the pipeline and re-evaluated its performance. The optimized version achieved a 17.5\% improvement in accuracy, along with significant gains in efficiency where concept annotation became over 10 times faster, and knowledge graph expansion over 100 times faster. These improvements stem from enhanced text extraction that uses sentence boundary detection and filtering of non-content text, the integration of a preprocessed Wikipedia dump database with precomputed embeddings reducing on-the-fly computation, and the use of a stronger transformer model all-mpnet-base-v2 for high quality embedding generation. Furthermore, the disambiguation module improved concept weighting, resulting in only highly relevant concepts being included in the EduKG. The updated worker-queue architecture also contributed to improved stability and reduced resource consumption, trading higher storage use for much lower CPU and memory demands. The pipeline is implemented within our MOOC platform, CourseMapper, to showcase its practical applicability in authentic educational settings. The final pipeline presents a state-of-the-art solution for automatic EduKG construction, with the flexibility to be adapted across various educational settings and use cases.

Despite these advances, some limitations remain. A considerable portion of runtime is spent saving expanded concepts in Neo4j, indicating a need for optimization in database interactions. The current data preprocessing step lacks support for incremental updates of the Wikipedia dump, requiring full reprocessing even for minor changes. Additionally, the text extraction module struggles with ligatures, hyphenated words, umlauts, and accents, and would benefit from more sophisticated layout analysis to better handle complex elements like tables and figures. The evaluation dataset (CCI) is also limited in coverage and outdated, constraining the benchmarking accuracy. A more representative dataset with learning materials actively used in practice and annotated by experts would provide a stronger evaluation framework. As future work, we plan to incorporate a human-in-the-loop mechanism, enabling domain experts to refine, validate, and enhance the automatically generated EduKGs before publishing it for the end users. This integration aims to strike a balance between automation and human expertise, ensuring high-quality EduKGs while significantly reducing the manual effort involved.
%
\begin{credits}
\subsubsection{\discintname}
The authors have no competing interests to declare that are
relevant to the content of this article. 
\end{credits}
%
%
%
%
\bibliographystyle{splncs04}  
\bibliography{references} 

\begin{thebibliography}{10}
\providecommand{\url}[1]{\texttt{#1}}
\providecommand{\urlprefix}{URL }
\providecommand{\doi}[1]{https://doi.org/#1}

\bibitem{info14100526}
Ain, Q.U., Chatti, M.A., Bakar, K.G.C., Joarder, S., Alatrash, R.: Automatic construction of educational knowledge graphs: A word embedding-based approach. Information  \textbf{14} (2023). \doi{10.3390/info14100526}

\bibitem{ain2022learning}
Ain, Q.U., Chatti, M.A., Joarder, S., Nassif, I., Wobiwo~Teda, B.S., Guesmi, M., Alatrash, R.: Learning channels to support interaction and collaboration in coursemapper. In: Proceedings of the 14th International Conference on Education Technology and Computers. pp. 252--260 (2022)

\bibitem{QURATlak}
Ain, Q.U., Chatti, M.A., Meteng~Kamdem, P.A., Alatrash, R., Joarder, S., Siepmann, C.: Learner modeling and recommendation of learning resources using personal knowledge graphs. In: Proceedings of the 14th Learning Analytics and Knowledge Conference. p. 273–283 (2024). \doi{10.1145/3636555.3636881}

\bibitem{ain2025top}
Ain, Q.U., Chatti, M.A., Shakhshir, A., Qussa, J., Alatrash, R., Joarder, S.: Top-down vs. bottom-up approaches for automatic educational knowledge graph construction in coursemapper. arXiv preprint arXiv:2505.10069  (2025)

\bibitem{umaprawaa}
Alatrash, R., Chatti, M.A., Ain, Q.U., Joarder, S.: Transparent learner knowledge state modeling using personal knowledge graphs and graph neural networks. In: Adjunct Proceedings of the 32nd ACM Conference on User Modeling, Adaptation and Personalization. p. 591–596 (2024). \doi{10.1145/3631700.3665230}

\bibitem{semeval}
Augenstein, I., Das, M., Riedel, S., Vikraman, L., McCallum, A.: Semeval 2017 task 10: Scienceie-extracting keyphrases and relations from scientific publications. arXiv preprint arXiv:1704.02853  (2017)

\bibitem{bast_2017_a}
Bast, H., Korzen, C.: A benchmark and evaluation for text extraction from pdf (06 2017). \doi{10.1109/JCDL.2017.7991564}

\bibitem{bmc_2018_bmc}
BMC: Bmc, research in progress (2018), \url{https://www.biomedcentral.com/}

\bibitem{chen2018knowedu}
Chen, P., Lu, Y., Zheng, V.W., Chen, X., Yang, B.: Knowedu: A system to construct knowledge graph for education. Ieee Access  \textbf{6},  31553--31563 (2018)

\bibitem{engelbrecht2024teaching}
Engelbrecht, E., Ilkou, E., Abu-Rasheed, H., Chaves-Fraga, D., Jim{\'e}nez-Ruiz, E., Labra-Gayo, J.E.: Teaching knowledge graph for knowledge graphs education. Semantic Web: interoperability, usability, applicability  (2024)

\bibitem{fettach2022knowledge}
Fettach, Y., Ghogho, M., Benatallah, B.: Knowledge graphs in education and employability: A survey on applications and techniques. IEEE Access  \textbf{10} (2022)

\bibitem{Gao2019}
Gao, J., Li, X., Xu, Y.E., Sisman, B., Dong, X.L., Yang, J.: Efficient knowledge graph accuracy evaluation. arXiv preprint arXiv:1907.09657  (2019)

\bibitem{keybert}
Grootendorst, M.: Keybert: Minimal keyword extraction with bert. \url{https://maartengr.github.io/KeyBERT} (2020)

\bibitem{hogan2021knowledge}
Hogan, A., Blomqvist, E., Cochez, M., d’Amato, C., Melo, G.D., Gutierrez, C., Kirrane, S., Gayo, J.E.L., Navigli, R., Neumaier, S., et~al.: Knowledge graphs. ACM Computing Surveys (CSUR)  \textbf{54}(4),  1--37 (2021)

\bibitem{inspec}
Hulth, A.: Improved automatic keyword extraction given more linguistic knowledge. In: Proceedings of the 2003 conference on Empirical methods in natural language processing. pp. 216--223 (2003)

\bibitem{jhajj2024educational}
Jhajj, G., Zhang, X., Gustafson, J.R., Lin, F., Lin, M.P.C.: Educational knowledge graph creation and augmentation via llms. In: International Conference on Intelligent Tutoring Systems. pp. 292--304. Springer (2024)

\bibitem{Manrique2018knowledge}
Manrique, R., Gr{\'e}visse, C., Mari{\~n}o, O., Rothkugel, S.: Knowledge graph-based core concept identification in learning resources. In: Joint International Semantic Technology Conference. pp. 36--51. Springer (2018)

\bibitem{manrique2018weighting}
Manrique, R., Marino, O.: Knowledge graph-based weighting strategies for a scholarly paper recommendation scenario. In: KaRS@ RecSys. pp.~5--8 (2018)

\bibitem{mendes2011dbpedia}
Mendes, P.N., Jakob, M., Garc{\'\i}a-Silva, A., Bizer, C.: Dbpedia spotlight: shedding light on the web of documents. In: Proceedings of the 7th international conference on semantic systems. pp.~1--8 (2011)

\bibitem{qu2024survey}
Qu, K., Li, K.C., Wong, B.T., Wu, M.M., Liu, M.: A survey of knowledge graph approaches and applications in education. Electronics  \textbf{13}(13), ~2537 (2024)

\bibitem{ramakrishnan2012layout}
Ramakrishnan, C., Patnia, A., Hovy, E., Burns, G.A.: Layout-aware text extraction from full-text pdf of scientific articles. Source code for biology and medicine  (2012)

\bibitem{reimers2019sentence}
Reimers, N., Gurevych, I.: Sentence-bert: Sentence embeddings using siamese bert-networks. arXiv preprint arXiv:1908.10084  (2019)

\bibitem{huggingface2023msmarco}
Reimers, N., Gurevych, I.: {msmarco-distilbert-base-tas-b}. \url{https://huggingface.co/sentence-transformers/msmarco-distilbert-base-tas-b} (2023)

\bibitem{pretrained}
Reimers, N., Gurevych, I.: Sentence-bert: Pretrained models (2023), \url{https://www.sbert.net/docs/sentence_transformer/pretrained_models.html}

\bibitem{mistral}
{TheBloke}: {TheBloke/Mistral-7B-Instruct-v0.1-GGUF}. \url{https://huggingface.co/TheBloke/Mistral-7B-Instruct-v0.1-GGUF} (2024)

\bibitem{tkaczyk2014grotoap2}
Tkaczyk, D., Szostek, P., Bolikowski, L.: Grotoap2-the methodology of creating a large ground truth dataset of scientific articles. D-Lib Magazine  \textbf{20}(11/12) (2014)

\bibitem{zhao2022edukg}
Zhao, B., Sun, J., Xu, B., Lu, X., Li, Y., Yu, J., Liu, M., Zhang, T., Chen, Q., Li, H., et~al.: Edukg: a heterogeneous sustainable k-12 educational knowledge graph. arXiv preprint arXiv:2210.12228  (2022)

\bibitem{2023surveykgs}
Zhong, L., Wu, J., Li, Q., Peng, H., Wu, X.: A comprehensive survey on automatic knowledge graph construction. ACM Comput. Surv.  (2023)

\bibitem{zhu2024llms}
Zhu, Y., Wang, X., Chen, J., Qiao, S., Ou, Y., Yao, Y., Deng, S., Chen, H., Zhang, N.: Llms for knowledge graph construction and reasoning: Recent capabilities and future opportunities. World Wide Web  \textbf{27}(5), ~58 (2024)

\end{thebibliography}
\end{document}